\documentclass[
aps,
prb,
groupedaddress,
superscriptaddress,
floatfix,
twocolumn,
notitlepage
]{revtex4-1}

\usepackage{graphicx}% Include figure files
\usepackage{amsmath,amssymb}
\usepackage{braket}
\usepackage{hyperref} % hypertext capabilities
\usepackage{subfigure}
\usepackage{bm} % bold math
\usepackage{mathtools} % dcases (uncompressed fractions)
\usepackage{float}

\def\Im{{\rm Im\,}}

\begin{document}

%\preprint{}

\title{Kagome Lattice from Exciton-Polariton Perspective}

\author{D. R. Gulevich} \email{d.r.gulevich@lmc.ifmo.ru}
\affiliation{ITMO University, St. Petersburg 197101, Russia}

\author{D. Yudin}
\affiliation{ITMO University, St. Petersburg 197101, Russia}
\affiliation{Division of Physics and Applied Physics, Nanyang Technological University 637371, Singapore}

\author{I. V. Iorsh}
\affiliation{ITMO University, St. Petersburg 197101, Russia}
\affiliation{Division of Physics and Applied Physics, Nanyang Technological University 637371, Singapore}

\author{I. A. Shelykh}
\affiliation{ITMO University, St. Petersburg 197101, Russia}
\affiliation{Division of Physics and Applied Physics, Nanyang Technological University 637371, Singapore}
\affiliation{Science Institute, University of Iceland, Dunhagi 3, IS-107, Reykjavik, Iceland}

\date{\today}

\begin{abstract} 
We study a system of microcavity pillars arranged into a kagome lattice. We show that polarization-dependent tunnel coupling of microcavity pillars leads to the emergence of the effective spin-orbit interaction consisting of the Dresselhaus and Rashba terms, similar to the case of polaritonic graphene studied earlier. Appearance of the effective spin-orbit interaction combined with the time-reversal symmetry-breaking resulting from the application of the magnetic field leads to the nontrivial topological properties of the Bloch bundles of polaritonic wavefunction. These are manifested in opening of the gap in the band structure and topological edge states localized on the boundary. 
Such states are analogs of the edge states arising in topological insulators.
Our study of polarization properties of the edge states clearly demonstrate that opening of the gap is associated with the band inversion in the region of the Dirac points of the Brillouin zone where the two bands corresponding to polaritons of opposite polarizations meet. 
For one particular type of boundary we observe a highly nonlinear energy dispersion of the edge state which makes polaritonic kagome lattice a promising system for observation of edge state solitons.
\end{abstract}

%\pacs{}
%\keywords{}                             

%\tableofcontents

%\maketitle
{\let\newpage\relax\maketitle}

%-----------------------------------------------------------------------------------------------
%-----------------------------------------------------------------------------------------------
\section{Introduction}
%-----------------------------------------------------------------------------------------------
%-----------------------------------------------------------------------------------------------

The modern theory of phase transitions originates from the suggestion of Landau that the transition from one state of matter to another must correspond to a spontaneous symmetry breaking~\cite{Landau}. This idea gives rise to the phenomenological theory of phase transitions which is formulated in terms of the order parameter. Quantum Hall effect (QHE)~\cite{Prange}, discovered more than 30 years ago, has posed some awkward questions regarding the original Landau approach. Indeed, the state of electrons in QHE might be treated as a  phase since the macroscopic observables such as the quantized Hall conductivity are not affected by smooth variations of material parameters. Still the transitions between different Hall plateaus are not related to violation of any underlying symmetry~\cite{Wen1989}.

The puzzle has been successfully resolved by introduction of the concept of topological order~\cite{Wen1990}. The quasiclassical dynamics of the wave packet propagating in periodic dissipative media turned out to be instrumental for the theoretical understanding of this type of order. In an attempt to explain the anomalous Hall effect in a ferromagnet as an intrinsic property of the band structure, Karplus and Luttinger \cite{Karplus-PR-1954} pointed out that the position operator in a periodic lattice fails to commute with itself. As a consequence the standard quasiclassical expression for the group velocity acquires an anomalous term, which is proportional to what is now called Berry curvature~\cite{Sundaram1999}. When integrated over a Brillouin zone of a two-dimensional lattice the Berry curvature gives topological invariant known as Chern number. The system possessing a non-zero topological invariant may be called topological matter~\cite{Wen1995}.

Observation of the spin Hall effect in a class of semiconductors with strong spin-orbit coupling~\cite{Kane-PRL-2005,Bernevig-Science-2006,Konig-Science-2007,Hsieh-Nature-2008} has revived an interest to topological phases of matter. A term topological insulator has been coined to describe a system which behaves like an insulator in the bulk but has conducting surface. The conduction is due to the presence of the edge states which possess remarkable property of topological protection: 
%It is predicted that electrons traveling along the surface are protected against back-scattering and preserve their phase coherence over long distances despite the presence of impurities, interactions, and external fields.
%It is predicted that 
electrons traveling along the surface of a topological insulator are protected against back-scattering on impurities.
%, interactions, and external fields.

\begin{figure}
\begin{center}
\includegraphics[width=3.5in]{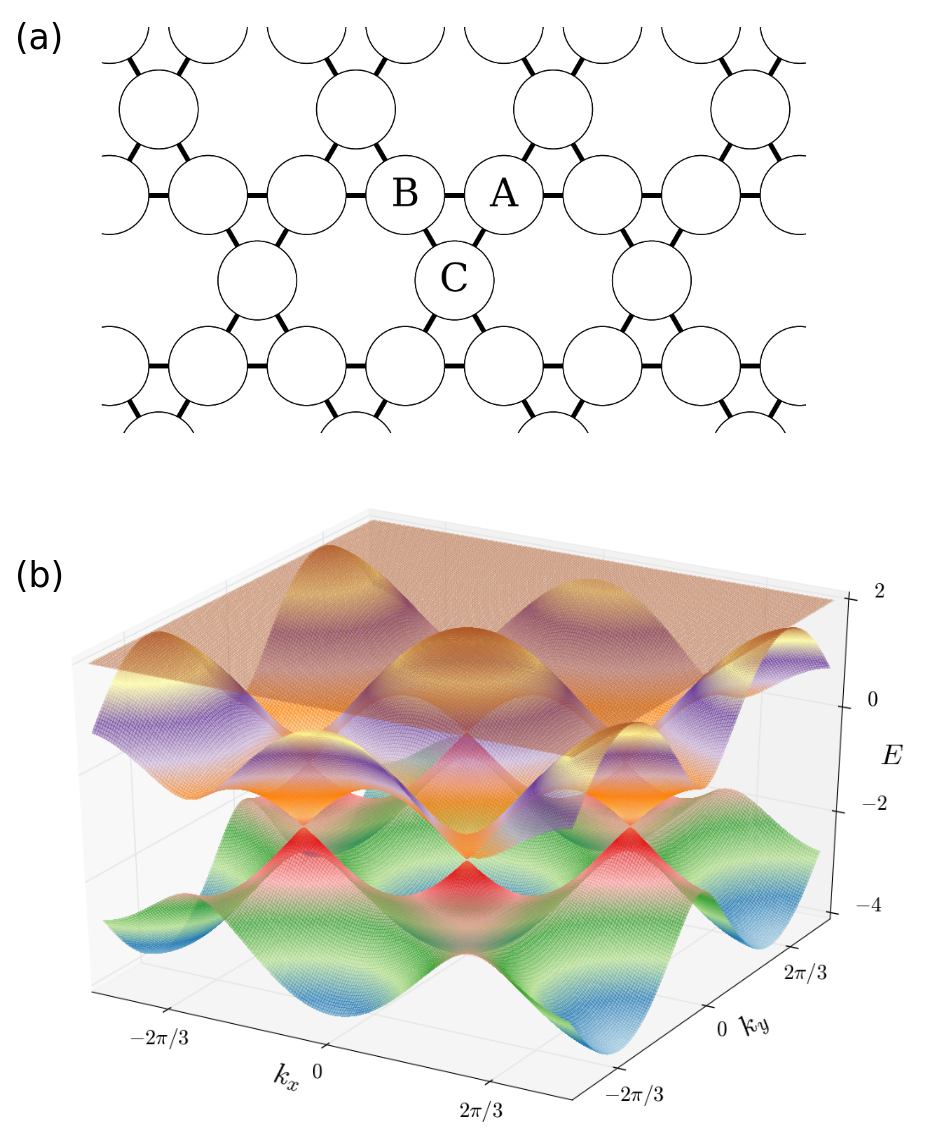}
\caption{\label{fig:kagome} 
(a) Sketch of two-dimensional kagome lattice formed by coupled microcavity pillars. Labels $A$, $B$, $C$ mark three inequivalent sites of the unit cell.	
(b) Band structure of the kagome lattice in the absence of magnetic field ($\Omega=0$) and TE-TM splitting ($\delta J=0$).
}
\end{center}
\end{figure}

It has been recently realized that topologically non-trivial phases may arise not only in the condensed matter electronic systems but also in photonic structures~\cite{Haldane2008, Raghu-2008, Wang-PRL-2008, Wang-2009, Hafezi-2011, Fang-2012, Rechtsman-2013, Khanikaev2013} (see, also, review~\cite{Lu-2014} and references therein). The artificial gauge fields in photonic structures can be realized in a system with periodic variation of dielectric permittivity. Large wavelength and coherence length of photons makes it simpler to realize diverse non-trivial landscapes of photonic nanostructures as compared to the electronic counterparts. At the same time, magnetic field control of the topological properties in photonic structures is precluded at optical frequencies because of the lack of natural magnetic materials in the optical frequency range (certain realizations of ferromagnetic photonic crystals for microwave region have been proposed and experimentally realized in Ref.~\onlinecite{Wang-2009}).

Thus, it would be useful to have a physical system which allows flexible control over topological properties via the external magnetic field and the landscape of periodic potential. Exciton-polaritons, quasi-particles originating due to the strong-light matter coupling between the quantum well excitons and cavity photons~\cite{Carusotto-Ciuti-RMP-2013} are good candidates. Exciton-polaritons attract significant interest for the last two decades because of the opportunities they offer for the observation of a wide class of quantum collective phenomena ranging from BEC and superfluidity~\cite{Kasprzak2006, Balili-2007, Amo-NP-2009, Amo-Nature-2009, Deng-RMP-2010, Sanvitto-2011, Nelsen-2013} to polaritonic lasing \cite{Imamoglu-PRA-1996, Deng-2003, Christopoulos-2007, Tsintzos-2008, Bhattacharya-2013, Schneider-2013, Kochereshko-2016}. Due to their hybrid light-matter nature, exciton-polaritons can be controlled with external magnetic field by inducing the Zeeman splitting in the excitonic component and structure patterning which forms trapping potential for the photonic component. 
Various techniques have been proposed for creation of the potential landscapes for polaritons. They include surface acoustic waves~\cite{de-Lima-2006, Cerda-Mendez-2012, Mendez-PRL-2010}, metal deposition~\cite{Lai-Nature-2007, Kim-2011, Masumoto-NJP-2012, Kim-2013, Kusudo-2013}, 
%beam litography and etching of the sample producing 
patterning of the planar structure and fabrication of arrays of coupled micropillars~\cite{Galbiati-PRL-2012, Abbarchi-NP-2013, Jacqmin-PRL-2014, Milicevic-2015}.
Moreover, polaritonic systems exhibit analog of spin-orbit interaction~\cite{Kavokin-PRL-2005, Leyder-NP-2007, Sala-PRX-2015} stemming from the splitting between TE and TM photonic modes~\cite{Kuther-PRB-1998, Panzarini-PRB-1999, Dasbach-PRB-2005}.

Recently there have been several proposals to realize two-dimensional topological phases of polaritons~\cite{Karzig-PRX-2015, Bardyn-PRB-2015, Nalitov-Z, Yi-PRB-2016}. 
For this, one needs to identify a system with
%with spin-orbit coupling and 
band structure characterized by bands touching at the point degeneracies such as Dirac cones.
% which split as soon as the time-reversal symmetry is broken. 
%Recently there have been several proposals to simulate topological phases of condensed matter systems with  polaritons~\cite{Karzig-PRX-2015, Bardyn-PRB-2015, Nalitov-Z}. 
%For this, one needs to identify systems with time-reversal and inversion symmetries and band structure characterized by the touching bands which split as soon as the time-reversal is broken. 
%For this, one needs to identify a system with band structure characterized by the touching bands which split as soon as the time-reversal is broken. 
%In Refs.~\onlinecite{Bardyn-PRB-2015} and~\onlinecite{Nalitov-Z} 
%In Ref.~\onlinecite{Nalitov-Z} polarization-dependent tunneling~\cite{Vasco-APL-2011} 
%due to TE-TM splitting of the photonic component 
%has been used to induce the effective spin-orbit coupling of polaritons in a honeycomb lattice~\cite{Milicevich}. 
%A promising way to implement the spin-orbit coupling in a lattice is by use of polarization-dependent tunneling~\cite{Vasco-APL-2011} as done in the recent study of honeycomb lattice~\cite{Nalitov-Z}.
To induce the spin-orbit coupling one may rely on the polarization-dependent tunneling~\cite{Vasco-APL-2011} 
%which has been shown to be of u as in the recent study of honeycomb lattice
proven to be useful in the context of polaritonic honeycomb lattice~\cite{Nalitov-graphene,Nalitov-Z}.
Breaking the time-reversal symmetry of a system by application of the magnetic field opens a gap in the band structure 
%As soon as the opened gap is 
characterized by non-zero Chern numbers for the bands below and above the gap.
%, this indicates a nontrivial topological phase.
The boundary between topological phases with different Chern numbers 
%(or other topological invariants) 
must support edge modes which are topologically protected against variations of the material parameters unless the band gap in the bulk of the material is collapsed. Such modes are analogs of the edge states arising on a surface of two-dimensional topological insulators and therefore polaritonic systems can play a useful role of a simulator to study topological order in a well-controlled experimental environment. 
%A promising way to realize topological phase is by inducing the spin-orbit coupling via the use of polarization-dependent tunneling~\cite{Vasco-APL-2011} as in the recent study of honeycomb lattice~\cite{Nalitov-Z}. 
In this paper we consider in details an example of 
a system of high experimental relevance where topological phase of exciton-polaritons may be realized: kagome lattice of microcavity pillars connected by tunnel coupling.

The paper is organized as follows. In Section II we introduce the model of polaritonic kagome lattice. We demonstrate the appearance of energy gaps when time-reversal and rotational symmetries are broken by magnetic field and TE-TM splitting. In Section III we study topological properties of the band structure for polaritons in kagome lattice. Section IV is devoted to study of edge modes which arise due to non-trivial topology of polaritonic bands, paying particular attention to their localization and polarization properties. Finally, we discuss the feasibility of  experimental observation of the predicted topological phases in polaritonic systems.

\setlength{\unitlength}{0.1in}
\begin{figure*}
\begin{center}	
\includegraphics[width=7in]{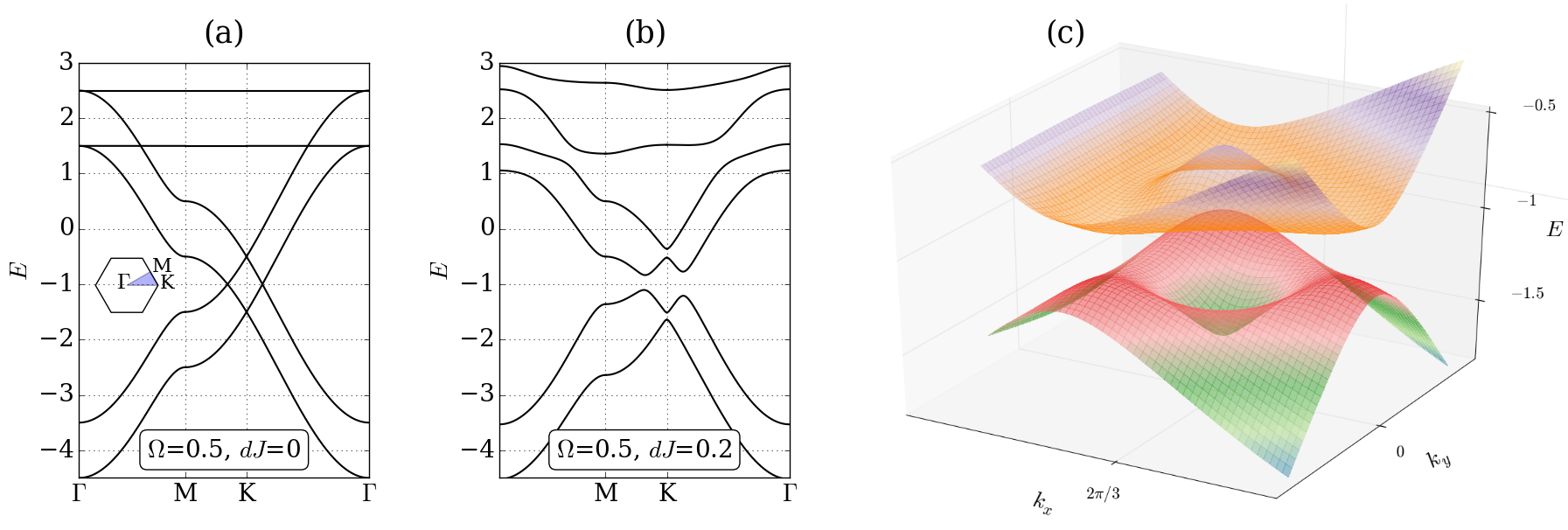}
\caption{\label{fig:bands} 
Mechanism of gap opening in polariton kagome lattice. (a) Kagome lattice band structure in presence of magnetic field $\Omega=0.5$ but no TE-TM splitting, $\delta J=0$. (b) In presence of magnetic field $\Omega=0.5$ and finite TE-TM splitting $\delta J=0.2$. (c) Zoom of the band structure at $\Omega=0.5$, $\delta J=0.2$ in the vicinity of one of the Dirac point $\mathbf{K}=(2\pi/3,0)$.
}
\end{center}
\end{figure*}

%-----------------------------------------------------------------------------------------------
%-----------------------------------------------------------------------------------------------
\section{Band Structure of Polaritonic Kagome Lattice}
%-----------------------------------------------------------------------------------------------
%-----------------------------------------------------------------------------------------------
We consider exciton-polariton microcavity pillars connected by tunnel coupling and arranged into a kagome lattice as shown in Fig.~\ref{fig:kagome}a.
Kagome lattice~\cite{kagome-PhysToday}, has a high degree of frustration compared to other 2D lattices, e.g. triangular and square lattices. It has been recently proposed as a candidate for magnetic material possessing  quasiparticles with topological properties~\cite{Pereiro-2014,Chisnell-PRL-2015}. In the tight-binding approximation with nearest-neighbor coupling $-J<0$ the highest energy band of kagome lattice band structure is completely flat while the lower bands touch at Dirac points in the corners of hexagonal Brillouin zone, see Fig.~\ref{fig:kagome}.
 
When TE-TM splitting is present in the junction connecting two microcavity pillars, the linear polarization modes in neighboring pillars experience different tunnel barriers~\cite{Vasco-APL-2011}. 
If $|L_j\rangle$, $|T_j\rangle$ are states of linear polarizations directed along and transverse to the direction connecting the centers of two neighbouring pillars~$j=1,2$,
then
$\braket{L_1|\hat V|L_2} = -J - \delta J, \quad \braket{T_1|\hat V|T_2} = -J + \delta J$
where $\hat{V}$ is the tunneling operator which links neighbouring pillars and ${2\,\delta J}$ is the difference in tunnel couplings of linearly polarized exciton-polaritons arising due to the TE-TM splitting. Such polarization-dependent tunneling has been shown to be responsible for appearance of effective spin-orbit interaction in polaritonic benzene molecule~\cite{Sala-PRX-2015} and polaritonic graphene~\cite{Jacqmin-PRL-2014, Kusudo-2013, Milicevic-2015,Nalitov-graphene,Nalitov-Z}. We will show that the polarization-dependent tunneling results in Rashba and Dresselhaus spin-orbit interaction terms in polaritonic kagome lattice and leads to opening of the gap in the vicinity of the Dirac points when time-reversal symmetry of the system is broken by external magnetic field. 

In the tight-binding approximation and with account of the polarization dependent coupling, the Hamiltonian for polaritons in the basis of circular polarization reads
\begin{equation}
\begin{split}
\hat{H}=
\Omega \sum_{i,\sigma=\pm}\sigma\,\hat{a}_{i,\sigma}^\dagger \hat{a}_{i,\sigma}
- J \sum_{\langle ij\rangle,\sigma=\pm} \left(\hat{a}_{i,\sigma}^\dagger \hat{a}_{j,\sigma}  + h.c.\right)\\ 
-  \delta J  \sum_{\langle ij\rangle} \left(
 e^{-2i\varphi_{ij}} \hat{a}_{i,+}^\dagger \hat{a}_{j,-} 
+ e^{2i\varphi_{ij}} \hat{a}_{i,-}^\dagger \hat{a}_{j,+} + h.c. \right)
\end{split}
\label{tb}
\end{equation}
Here, the summation $\langle ij\rangle$ is over nearest neighbors, operators $\hat{a}_{i,\sigma}^\dagger$ ($\hat{a}_{i,\sigma}$) create (annihilate) exciton-polariton of circular polarization $\sigma$ at site $i$ of the kagome lattice,
angles 
$\varphi_{ij}$ specify directions of vectors connecting the neighboring sites. The first term in~\eqref{tb} describes the Zeeman energy splitting ($2\Omega$) of the circular polarized components, the second term stands for the nearest neighbor hopping and the third term is the polarization-dependent coupling of cross-polarized polaritons in neighboring pillars.

The eigenstates of the tight-binding Hamiltonian~\eqref{tb} can be searched in the form of linear combination of the Bloch wavefunctions associated with
sublattices $A$, $B$ and~$C$ (see Fig.~\ref{fig:kagome}a),
\begin{equation}
\ket{\psi_{\mathbf{k}}} = \sum_{\sigma} \left( A_{\mathbf{k}}^{\sigma} \ket{\psi_{\mathbf{k}}^{A\sigma}}
+ B_{\mathbf{k}}^{\sigma} \ket{\psi_{\mathbf{k}}^{B\sigma}}
+ C_{\mathbf{k}}^{\sigma} \ket{\psi_{\mathbf{k}}^{C\sigma}} \right).
\end{equation}
Here, index $\sigma$ runs over the two circular polarizations and the Bloch wavefunctions $\ket{\psi_{\mathbf{k}}^{L\sigma}}$  with  $L=A,B,C$, 
are linear superpositions of states $\ket{\phi_j^{L\sigma}}$ localized on site~$j$,
\begin{equation}
\ket{\psi_{\mathbf{k}}^{L\sigma}} = \sum_{j} e^{- i \mathbf{k}\cdot \mathbf{R}_j^L} \ket{\phi_j^{L\sigma}}.
\label{anzatz}
\end{equation}
In the basis $\{\ket{\psi_{\mathbf{k}}^{A+}}$, $\ket{\psi_{\mathbf{k}}^{A-}}$, $\ket{\psi_{\mathbf{k}}^{B+}}$, $\ket{\psi_{\mathbf{k}}^{B-}}$, $\ket{\psi_{\mathbf{k}}^{C+}}$, $\ket{\psi_{\mathbf{k}}^{C-}}\}$
the Hamiltonian reads
\begin{equation}
\hat H_{\mathbf{k}} = 
\begin{pmatrix}
\Omega\hat\sigma_z & \hat F^{AB}_\mathbf{k} & \hat F^{AC}_\mathbf{k} \\[2pt]
\hat F^{AB}_\mathbf{k} & \Omega\hat\sigma_z & \hat F^{BC}_\mathbf{k} \\[2pt]
\hat F^{AC}_\mathbf{k} & \hat F^{BC}_\mathbf{k} & \Omega\hat\sigma_z
\end{pmatrix},
\label{Hk}
\end{equation}
where $\hat\sigma_z$ is the Pauli matrix and matrices $\hat F_\mathbf{k}^\mathbf{d}$ for $\mathbf{d}=AB,AC,BC$, are
\begin{equation}
\hat F_\mathbf{k}^{\mathbf{d}} = 
-2 \begin{pmatrix}
J \cos \mathbf{k}\cdot\mathbf{d} & \delta J e^{-2i\varphi_\mathbf{d}} \cos \mathbf{k}\cdot\mathbf{d}
\\[2pt]
\delta J e^{2i\varphi_\mathbf{d}} \cos \mathbf{k}\cdot\mathbf{d} & J \cos \mathbf{k}\cdot\mathbf{d}
\end{pmatrix}.
\end{equation}
The eigenstates and eigenenergies are defined by the stationary Schr\"{o}dinger equation
\begin{equation}
\hat H_{\mathbf{k}} u_{\mathbf{k},m} = E_{\mathbf{k},m} u_{\mathbf{k},m}.
\label{stSch}
\end{equation}
In what follows we will use normalized units by setting the characteristic length $|\mathbf{d}|$ and energy $J$ to unity. In the absence of the magnetic field $\Omega=0$ and TE-TM splitting $\delta J=0$ the arising band structure coincides with the standard band structure of kagome lattice shown in Fig.~\ref{fig:kagome}b. 
The presence of TE-TM splitting ($\delta J\neq 0$) breaks the rotational symmetry while
application of the magnetic field ($\Omega\neq 0$) breaks the time-reversal symmetry of the system.
This leads to the opening of a band gap at the Dirac points of the Brillouin zone, see Fig.~\ref{fig:bands}. The band structure arising from the gap opening in the vicinity of the Dirac point is shown in Fig.~\ref{fig:bands}c.

We analyze the Hamiltonian~\eqref{Hk} by performing decomposition in the vicinity of Dirac point $\mathbf{K}=(2\pi/3,0)$ followed by projection to the subspace formed by four lowest energy eigenstates calculated at $\delta J=0$, $\Omega=0$~(see details in Appendix). 
%Introducing operators $\hat\tau_i$ and $\hat\sigma_i$ acting on sublattice and polarization degrees of freedom, respectively, 
We introduce operator $\hat\sigma_i$ acting on polarization degrees of freedom
and operator $\hat\tau_i$ acting on two orthogonal sublattice modes (defined by Eq.~\eqref{v-basis} in Appendix).
Dropping an unimportant additive constant we get the effective Hamiltonian,
\begin{equation}
\hat H_{\rm eff} = \hat H_0 + \hat H_\mathbf{q},
\label{HDirac}
\end{equation}
where $\mathbf{q}=\mathbf{k}-\mathbf{K}$ and we explicitly separated ${\mathbf{q}\text{-independent}}$ 
\begin{equation}
\hat H_0 = \Omega \; \hat\sigma_z  + \delta J (\hat\tau_y \hat\sigma_x + \hat\tau_x \hat\sigma_y),
\label{H0}
\end{equation}
and $\mathbf{q}$-dependent part,
\begin{equation}
\hat H_\mathbf{q} = \sqrt{3}  ( \hat\tau_y q_x  - \hat\tau_x q_y ) + \alpha\,\hat\tau_y \,\pmb{\sigma}\cdot\pmb{q}\, + \hat\tau_x \hat H_R\, + \hat H_D,
\end{equation}
containing Rashba
\begin{equation}
\hat H_R = \alpha \left(\hat\sigma_x q_y - \hat\sigma_y q_x \right),
\end{equation}
and Dresselhaus
\begin{equation}
\hat H_D = \alpha \left(-\hat\sigma_x q_x + \hat\sigma_y q_y\right)
\end{equation}
spin-orbit interaction terms with coupling strength $\alpha=\delta J\sqrt{3}/2$. 
At the very Dirac point ($\mathbf{q}=0$) the four eigenenergies of the Hamiltonian~\eqref{HDirac} are those of the Hamiltonian~\eqref{H0}: $E=\pm \Omega$, $\pm \sqrt{4\delta J^2 + \Omega^2}$ measured relative to the energy of the Dirac point at $\delta J=0$, $\Omega=0$. 
%At small values of $\mathbf{q}$, eigenenergies of the Hamiltonian~\eqref{HDirac} are a good approximation to those of the full Hamiltonian~\eqref{Hk}, see Figs.~\ref{fig:bands}a-c for the comparison.

\setlength{\unitlength}{0.1in}
\begin{figure}
	\begin{center}
		\begin{picture}(35,20)
		\put(0,0){\includegraphics[width=3.4in]{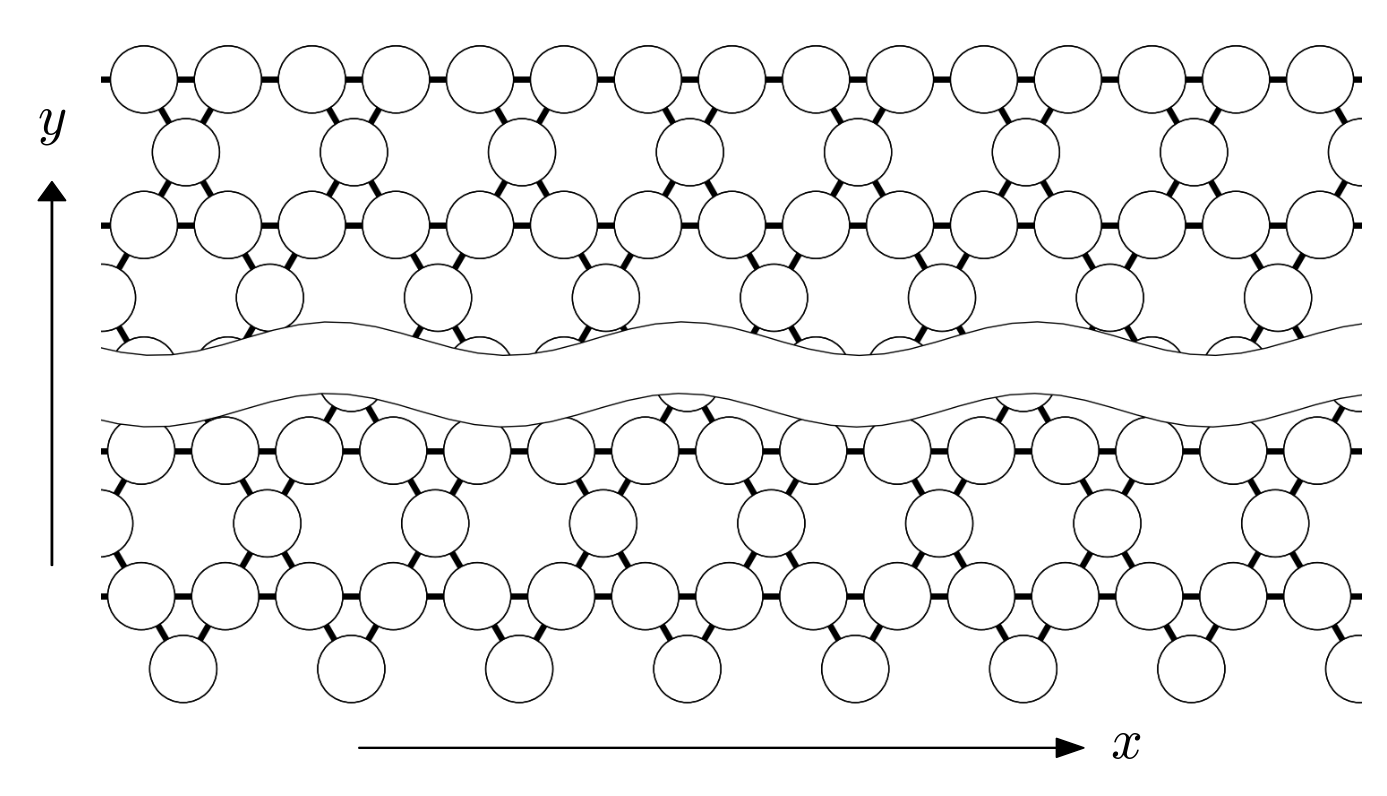}}
		\end{picture}
		\caption{\label{fig:strip} 
			A strip of kagome lattice. The strip is infinite along $x$ but has a finite extent along $y$ axis. The upper and lower boundaries are cut differently to study the effect of boundaries on the propagation of the topological edge modes.
		}
	\end{center}
\end{figure}

\begin{figure*}
\begin{center}
\includegraphics[width=7.2in]{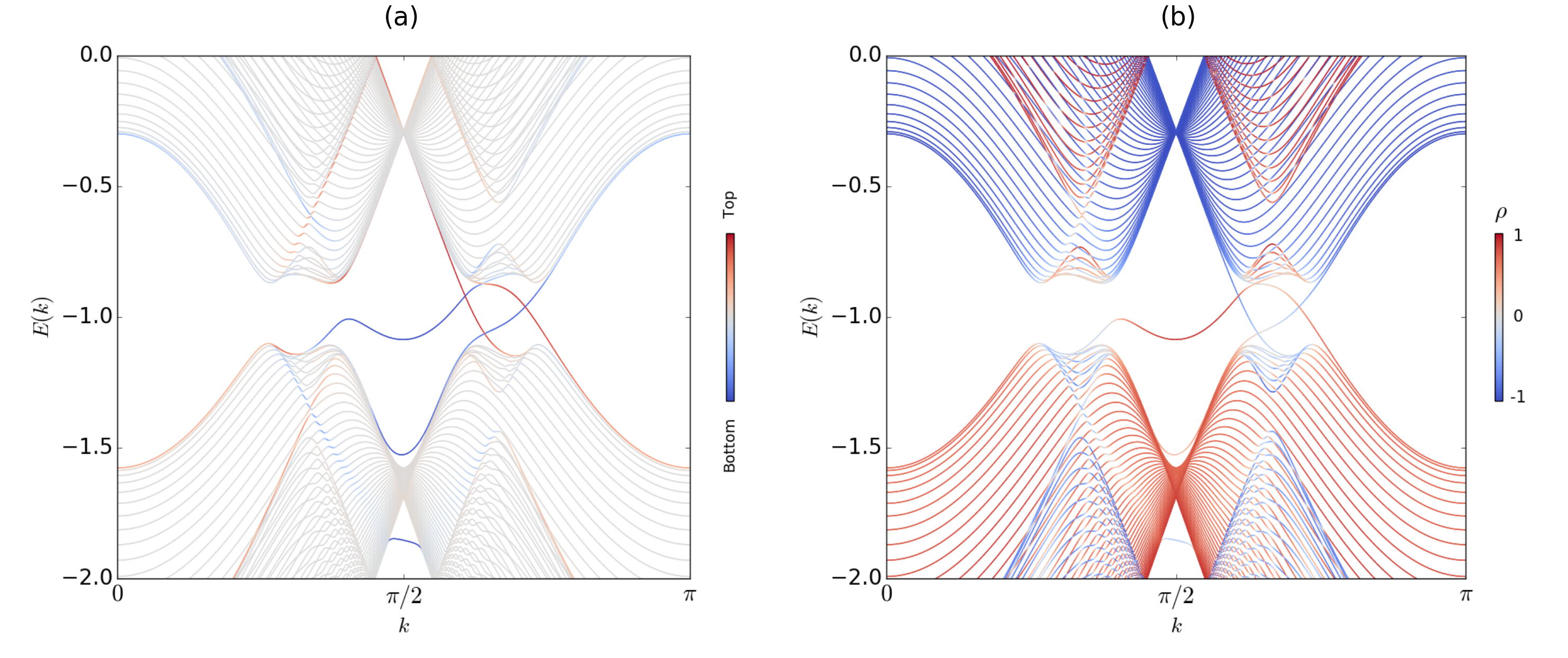}
\caption{\label{fig:disp-strip} 
(Color online) Band structure for the strip of kagome lattice shown on Fig.~\ref{fig:strip} at $\delta J=0.15$, $\Omega=0.3$. For convenience, the shifted Brillouin zone is used for better representation of the edge state dispersions. The color scale in (a) encodes localization of the eigenstates defined by the formula~\eqref{loc} where linear weight function was used. Red and blue color correspond to the states localized on the upper and lower boundary of the strip, respectively, see Fig.~\ref{fig:strip}. (b) represents the polarization degree given by the Eq.~\eqref{pol}. Band inversion causing the non-trivial topological twist can be seen in the areas above and below the gap where the red and blue lines mix. 
}
\end{center}
\end{figure*}

%-----------------------------------------------------------------------------------------------
%-----------------------------------------------------------------------------------------------
\section{Topology of polaritonic bands}
%-----------------------------------------------------------------------------------------------
%-----------------------------------------------------------------------------------------------

Topological ideas have become ubiquitous in condensed matter physics \cite{Schwarz,Monastyrsky,Nakahara} over the past few decades since the seminal work of Berry \cite{Berry} in which he demonstrated that the wave function of any quantum system gains an extra phase, Berry phase, (also called Pancharatnam-Berry phase, see the preceding work~\cite{Pancharatnam}) during adiabatic evolution around a closed path in momentum space. One of the most striking consequences of the Berry phase is a dramatic modification of the quasiclassical equation of motion 
for electron in a lattice~\cite{Sundaram-PRB-1999}.
%in a quasi-uniform field which is homogeneous on the scale of a unit cell. 
%The Berry curvature is analogous to magnetic field and contributes the term known as anomalous velocity to the equation of motion. In this respect, the Berry curvature restores the complete symmetry between real and reciprocal spaces with anomalous velocity term representing the current flow within the unit cell as the charge is redistributed between the orbitals.

It turns out that the appearance of Berry phase has a profound impact on properties of crystalline solids~\cite{Xiao-RMP-2010}.
% (see, e.g. Ref.~\onlinecite{Xiao-RMP-2010}). 
For non-interacting systems the eigenstates respect the periodicity of the Hamiltonian, so that the corresponding Brillouin zone can be considered as a parameter space of the Hamiltonian determined by the quasimomentum~$\mathbf{k}$. Quasimomenta which differ by a reciprocal lattice vector are to be identified. The Brillouin zone thus has topology of a torus. During evolution in quasimomentum space the underlying Bloch state picks up a phase shift which appears to be gauge-independent if the trajectory is closed.

To analyze topology of polaritonic bands it is enough to study the bands either below or above the gap of interest. This is due to the fact that topology of the whole Bloch bundle is trivial (see, e.g. review in Ref.~\onlinecite{Fruchart-review-2013}) while either of its halves reflects topology of the other. Chern number $C$ for the bands below the gap $E_g$ is given by 
\begin{equation}
C=\sum_{E_m<E_g} C_m = \frac{1}{2\pi} \sum_{E_m<E_g} \iint_{BZ} B_{\mathbf{k},m} d^2\mathbf{k},
\end{equation}
where the Berry curvature for the $m$th band is
\begin{equation}
B_{\mathbf{k},m} = -2\,\Im \Big\langle \frac{\partial}{\partial k_x} u_{\mathbf{k},m} , \frac{\partial}{\partial k_y} u_{\mathbf{k},m} \Big\rangle.
\end{equation}
%Due to the vanishing total Berry curvature~\cite{Fruchart-Ref22,Fruchart-Ref23} we have $C_{above}=-C_{below}$.
Integrating the Berry curvature over the Brillouin zone for the bands below/above the gap at $E_g=-1$ yields the Chern numbers $C=\pm 2$. According to the bulk-boundary correspondence, a non-zero difference in Chern numbers of the bands separated by the gap necessarily leads to existence of gapless edge states propagating along the boundary.
Such edge states are analogs of electronic edge modes of arising in topological insulators.

% While the electronic band structure of a topological insulator is gapped in the bulk, it allows traveling of gapless edge states that are topologically protected against inhomogeneities and variation of the material parameters.

%-----------------------------------------------------------------------------------------------
%-----------------------------------------------------------------------------------------------
\section{Polaritonic Edge states}
%-----------------------------------------------------------------------------------------------
%-----------------------------------------------------------------------------------------------

To study the edge states indicated by the non-trivial topology of polaritonic bands we consider a strip of kagome lattice which is infinite along $x$ and has a finite extent of $N_y=30$ unit cells along $y$, see Fig.~\ref{fig:strip}. We intentionally choose different boundary conditions at the upper and lower edge in order to study the effect of the boundaries on the dispersions of the edge modes.
The dispersion of the strip modes is presented in Fig.~\ref{fig:disp-strip}a as a function of momentum $k$ along the $x$ direction. We use the shifted Brillouin zone with $k$ in the range $[0,\pi)$ for a better display of the edge state dispersions. In accordance with calculation of the Chern numbers there emerge four edge states whose energies lie inside the gap.

To study localization of the eigenstates on the boundaries we define a quantity describing localization of a state $u_{k,m}^\sigma$ by forming a convolution with the weight function $p(y)$,
\begin{equation}
\lambda_{m}(k) = \sum_{\sigma,j} p(y_j) |u_{k,m}^\sigma (y_j)|^2,
\label{loc}
\end{equation}
where $y_j$ are positions of the cells along the $y$ axis. To obtain results presented in the Fig.~\ref{fig:disp-strip}a we used a linear localization weight function $p(y)\sim y-\frac12 \Delta y$, where $\Delta y$ 
%$\Delta y=(N_y-\frac12)\sqrt{3}$ 
is the width of the strip. As seen from the Fig.~\ref{fig:disp-strip}a, there are four dispersion curves which connect topologically non-trivial polaritonic bands. These correspond to the topological edge states localized on the two boundaries. The influence of different boundary conditions can be traced by noticing the asymmetry of the dispersions with respect to inversion of the quasimomentum $k\to -k$ (i.e. with respect to $k=\pi/2$ point for the shifted Brillouin zone on Fig.~\ref{fig:disp-strip}).

A peculiar feature arising at the chosen set of parameters ($\delta J =0.15$ and $\Omega=0.3$) is that the dispersion of the edge state propagating along the lower boundary in Fig.~\ref{fig:strip} is highly bent and forms well defined minimum and maximum. In the vicinity of maximum (minimum) the effective mass of the edge state is negative (positive) and suggests favorable conditions for existence of localized nonlinear edge excitations in the form of dark and bright solitons. Owing to such dispersion relation, polariton kagome lattice may be in advantage compared to other polariton lattices in search for a system which allows propagation of edge state solitons.

We then study polarization properties of edge states. For the $m$-th band we define the degree of polarization,
\begin{equation}
\rho_m(k)=\frac{N_{k,m}^{+} -N_{k,m}^{-}}{N_{k,m}^{+} + N_{k,m}^{-}},
\quad
N_{k,m}^\sigma=\sum_j |u_{k,m}^\sigma(y_j)|^2
\label{pol}
\end{equation}
The band diagram colored by polarization properties allows to see clearly the band inversion in the areas of the energy dispersion where the red and blue curves corresponding to the two polarizations overlap, see Fig.~\ref{fig:disp-strip}b. 
Although, such band inversion is not a sufficient condition for non-trivial topological twist to arise (see, e.g. Ref.~\onlinecite{Volkov-2016})
, it is an important mechanism for formation of non-trivial topological phase in electronic systems~\cite{Bernevig-Science-2006, Li-marginal}.

In order to observe experimentally the predicted effects in polaritonic kagome lattice one needs to open the gap in the band structure which exceeds the measured linewidth of the system.
As an example, for realistic parameters $J=1$ meV, $\Omega = 100\;\rm \mu eV$ and $\delta J=100\;\rm \mu eV$ in physical units the opened gap will be about $150\;\rm \mu eV$. In realistic systems the emission linewidth will limited from below by disorder broadening and lifetime of exciton-polariton. The last limitation may in principle be removed by the use of indirect excitons with sufficiently longer lifetimes as proposed in Ref.~\onlinecite{Bardyn-PRB-2015}. In order to further increase the gap larger values of magnetic field and TE-TM splitting are desirable. Large values of TE-TM splitting have been recently reported in open resonators, exceeding by a factor of three TE-TM spitting in monolithic microcavities~\cite{Duff-PRL-2015}. One more attractive alternative is presented by the use of exciton-polaritons in waveguides~\cite{Walker-APL-2013}.

%-----------------------------------------------------------------------------------------------
%-----------------------------------------------------------------------------------------------
\section{Conclusions}
%-----------------------------------------------------------------------------------------------
%-----------------------------------------------------------------------------------------------

To conclude, we analyzed topological properties of the system of microcavity pillars arranged into a kagome lattice and find that the effective spin-orbit interaction induced by the TE-TM splitting leads to opening of gap in the dispersion in presence of magnetic field. Analysis of the topology of polaritonic Bloch bands reveals non-zero Chern numbers and indicate the presence of topologically protected edge states.
% with energies lying inside the gap. 
Due to the highly nonlinear dispersion of edge states, kagome lattice may be of interest in studies of essentially nonlinear effects such as propagation of edge solitons in polaritonic lattices.

%Fig.~\ref{bs-Omega}a lies within the experimentally accessible range of the magnetic fields (for $J\sim 1\;\rm meV$, the half-energy Zeeman splitting $\Omega/J\sim 0.1$ corresponds to a magnetic field about 10 Tesla~\cite{Larionov-PRL-2010, Walker-PRL-2011, Fisher-PRL-2014, Sturm-PRB-2015}).

%-----------------------------------------------------------------------------------------------
%-----------------------------------------------------------------------------------------------
\begin{acknowledgments}
    We acknowledge support of the Ministry of Education and Science of the Russian Federation in the framework of Increase Competitiveness Program 5-100 and of Singaporean Ministry of Education under AcRF Tier 2 grant MOE2015-T2-1-055. D.Y. acknowledges support from RFBR project 16-32-60040. I.V.I. appreciates the support of the Ministry of Education and Science of the Russian Federation (Zadanie No. 3.1231.2014/K), Grant of the President of Russian Federation (MK-5220.2015.2) and RFBR project 16-32-60123. I.V.I. and I.A.S. thank the support from Rannis excellence grant 163082-051.
\end{acknowledgments}
%-----------------------------------------------------------------------------------------------
%-----------------------------------------------------------------------------------------------

%=============================================================================
%=============================================================================
% Following sections are appendices. 
% Use \appendix* if there is only one appendix.
%\appendix

%-----------------------------------------------------------------------------------------------
%-----------------------------------------------------------------------------------------------
\appendix*
\section{Projection onto the eigenspace around the Dirac point}
%-----------------------------------------------------------------------------------------------
%-----------------------------------------------------------------------------------------------

We follow Ref.~\onlinecite{projection} to perform projection of the Hamiltonian~\eqref{Hk} to the subspace formed by the eigenstates at the Dirac point. To get the effective Hamiltonian~\eqref{HDirac} we used the following set of four degenerate eigenstates of the Hamiltonian~\eqref{Hk} at $\mathbf{k}=\mathbf{K}$, $\Omega=0$, $\delta J=0$,
\begin{equation}
u_{\mathbf{K},i,\pm}=v_i\otimes s_{\pm}.
\label{basis}
\end{equation}
Here,
\begin{equation}
v_{1}=\frac{1}{\sqrt{3}}
\begin{pmatrix}
 \frac{1 + i\sqrt{3}}{2} \\  \frac{1 - i\sqrt{3}}{2} \\ 1
\end{pmatrix} ,\;
v_2=-\,\frac{i}{\sqrt{3}}
\begin{pmatrix}
 \frac{1 - i\sqrt{3}}{2} \\  \frac{1 + i\sqrt{3}}{2} \\ 1
\end{pmatrix}
\label{v-basis}
\end{equation}
are two lattice modes and 
\begin{equation}
s_+ = 
\begin{pmatrix}
1 \\ 0
\end{pmatrix},\quad
s_- = 
\begin{pmatrix}
0 \\ 1
\end{pmatrix}
\label{s-basis}
\end{equation}
correspond to the two circular polarization states. 
%Basis vectors~\eqref{basis} are four degenerate solutions to the stationary Schroedinger equation~\eqref{stSch} at $\mathbf{k}=\mathbf{K}$, $\Omega=0$, $\delta J=0$. 
Note, that the choice of lattice modes~\eqref{v-basis} is not unique due to the degeneracy at the Dirac point but defined up to an arbitrary rotation. We define $\hat\tau_i$ and $\hat\sigma_i$ operators as Pauli matrices acting in space spanned by the lattice modes~\eqref{v-basis} and polarization states~\eqref{s-basis}, respectively.

\end{document}